\documentstyle[12pt]{article}
\title{Effective potential for a covariantly constant gauge field in
curved spacetime}
\author{
\normalsize Emilio Elizalde$^{a,b,}$\thanks{E-mail:
eli@zeta.ecm.ub.es},
Sergei D. Odintsov$^{b,}$\thanks{On
leave of absence from Tomsk Pedagogical University,
634041 Tomsk, Russia. E-mail: odintsov@ecm.ub.es},
and August Romeo$^{a,}$\thanks{E-mail:
august@ceab.es}, \\
\normalsize $^a$Centre for Advanced Studies CEAB, CSIC,
Cam\'{\i} de Santa B\`arbara, 17300 Blanes \\
\normalsize $^b$Dept ECM and IFAE,
Fac. Physics, University of  Barcelona,
Diagonal 647, 08028 Barcelona
}
\date{}

\begin{document}

\maketitle

\begin{abstract}
We discuss the influence of gravitational effects on the stabilization
of the chromomagnetic vacuum. The one-loop effective potential for a
covariantly constant SU(2) gauge field in ${\bf S}^2 \times {\bf R}^2$
and ${\bf T}^2 \times {\bf R}^2$
is calculated. A possibility of curvature-induced phase transitions
between zero and nonzero chromomagnetic vacua is found ---what is also
confirmed through the calculation of the renormalization group (RG)
improved effective potential on constant-curvature spaces with small
curvature.  Numerical evaluation indicates that for some curvatures
the imaginary part of the effective potential disappears (gravitational
stabilization of the chromomagnetic vacuum occurs).
\end{abstract}

\section{Introduction}
The study of the effective Lagrangian for covariantly constant gauge
fields in nonabelian gauge theories showed the possibility of
existence for a nonzero minimum (the so-called chromomagnetic vacuum
\cite{Sa}). Unfortunately, such a state could not be the true ground
state of the theory due to the presence of unstable field modes which
lead to the appearance of an imaginary part in the effective Lagrangian
\cite{NO} (see also  \cite{ES}-\cite{FGS} and references
therein). Several proposals for stabilizing the vacuum may be
considered. For example, in the interesting paper \cite{AO} it has been
shown that a stable ground state in electroweak theory may be achieved
for a space-dependent magnetic field.

In the present letter, we investigate the effects of nonzero curvature
on the stabilization of the chromomagnetic vacuum for a pure gauge theory
in curved spacetime. Using zeta-regularization methods we calculate the
effective potential for an SU(2) covariantly constant gauge field in a
spacetime of the form ${\bf M}^2 \times {\bf R}^2$, where ${\bf M}^2$ is
a curved manifold ---here the case of
the two-dimensional sphere and two-dimensional torus will be considered---
 and ${\bf R}^2$ a flat plane. The nonzero
components of the background gauge field are them embedded into
${\bf R}^2$. An analysis of the phase structure from the effective
potential shows the chance of curvature-induced phase transitions
between zero and nonzero chromomagnetic ground states. Moreover, our
study indicates the possibility of removing the imaginary part by
curvature effects, thus stabilizing the chromomagnetic vacuum.
The possibility of curvature-induced phase transitions is also verified in
sect. 3 through the calculation of the RG-improved effective potential in
curved spacetime for small curvatures (where quasi-local expansions can
be used).

\section{Gauge field effective potential on an
${\bf S}^2 \times {\bf R}^2$ background}

Let us consider a pure SU($N$) gauge theory on an
${\bf S}^2 \times {\bf R}^2$ spacetime background with a non-zero
covariantly constant background gauge field. Like in flat spacetime,
the calculation of the one-loop effective action is quite straightforward,
giving the result
\begin{equation}
\Gamma^{(1)}[\overline{A}]=
{1 \over 2}{\rm Tr}\ \ln\left[ \overline{\Theta}_{\mu \nu}^{a b} \right]
-{\rm Tr}\ \ln\left[ (-\overline{D}^2)^{a b} \right]
\label{detdet}
\end{equation}
where
the operators present are
\begin{equation}
\begin{array}{lll}
\displaystyle\overline{\Theta}_{\mu \nu}^{a b}&=&\displaystyle
-g_{\mu\nu} 
\left( \overline{D}_{\lambda} \overline{D}^{\lambda} \right)^{a b}
+2i g (T^c)^{a b} \overline{F}_{\mu \nu}^c
+\left( 1 - {1/\xi} \right)
\left( \overline{D}_{\mu} \overline{D}_{\nu} \right)^{a b}
-\delta^{a b} R_{\mu \nu} ,
\hspace{5mm} \xi=1, \\
\displaystyle\overline{D}_{\mu}^{a b}&=&\displaystyle
\delta^{a b} \nabla_{\mu}-ig (T^c)^{ab} \overline{A}_{\mu}^c
\end{array}
\end{equation}
The $T^a$ matrices are the colour SU($N$) generators, which we will
suppose written in the adjoint representation.
Colour SU($N$) indices are inset in such a way that
\begin{equation}
\begin{array}{lll}
\displaystyle \overline{A}_{\mu}&=&
\displaystyle n^a \overline{A}_{\mu}^a , \\
\displaystyle \overline{F}_{\mu \nu}&=&
\displaystyle n^a \overline{F}_{\mu \nu}^a ,
\end{array}
\end{equation}
being $n$ a unit-norm vector along some given colour space direction.
Then
\begin{equation}
\overline{\Theta}_{\mu \nu}^{a b}=
 -g_{\mu\nu} \left[
\left( \nabla_{\lambda} -i g(n^c T^c)\overline{A}_{\lambda} \right)^2
\right]^{a b}
+ig (n^c T^c)^{a b} \overline{F}_{\mu \nu} -\delta^{a b} R_{\mu \nu} .
\label{defTheta}
\end{equation}
The Tr ln calculations in (\ref{detdet})
will be performed by the zeta function
method.
In view of that equation, the function whose derivative
at $s=0$ will provide the effective action is the
linear combination
\begin{equation}
-{1 \over 2} \zeta_{\overline{\Theta}/\mu^2}(s) 
+ \zeta_{-\overline{D}^2/\mu^2}(s),
\label{wholez}
\end{equation}
where the subscripts mean the operators to which each zeta function
is associated (for an introduction to zeta-regularization in full
detail, see the recent monographies \cite{EORBZ}).
Renormalization of gauge theories in curved spacetime
has been thoroughly discussed in refs. \cite{BO} (for a review
and introduction to quantum field theory in curved
spacetime, see \cite{BOS}).

We start by diagonalizing the colour-structure. Formally, one
identifies the eigenvalues of the colour matrices:
\begin{equation}
(n^c T^c) \ \mbox{with eigenvalues}  \ \nu^a, \, a=1, \dots , N^2-1
\end{equation}
Thus, we find the colour `eigenoperators'
\begin{equation}
\overline{\Theta}_{\mu \nu}^a = -g_{\mu\nu}
\left( \nabla_{\lambda} -i g \nu^a \overline{A}_{\lambda} \right)^2
+ig \nu^a \overline{F}_{\mu \nu} - R_{\mu \nu} , \
a=1, \dots , N^2-1
\end{equation}
Now, the selected type of background comes into play.
We choose, as a background gauge field,
\begin{equation}
\overline{A}_{\mu}= -{1 \over 2} \overline{F}_{\mu \nu} x^{\nu}
\end{equation}
with constant $\overline{F}_{\mu \nu}$.
It is covariantly constant, i.e.
it satisfies $\nabla_{\mu}\overline{A}^{\mu}=0$.

The spacetime in which we shall work is ${\bf S^2}\times {\bf R}^2$.
Ordering the coordinates in the way $( \theta, \varphi, x_1, x_2 )$,
we consider a purely colour-magnetic case for which
\begin{equation}
(\overline{F}_{\mu \nu})=\left(
\begin{array}{cccc}
0&0& 0&0 \\
0&0& 0&0 \\
0&0& 0&H \\
0&0&-H&0
\end{array}
\right)
\end{equation}
As a result,
\begin{equation}
\Delta^a \equiv
\left( \nabla_{\lambda} -i g \nu^a \overline{A}_{\lambda} \right)^2
=\nabla^2 +ig \nu^a H (x_1 \partial_2 - x_2 \partial_1)
-{g^2 (\nu^a)^2 \over 4} H^2 (x_1^2+x_2^2) .
\end{equation}
Since our spacetime is ${\bf S^2}\times {\bf R}^2$,
$\displaystyle
\nabla^2= \nabla^2_{{\bf S}^2} + \partial_1^2 + \partial_2^2
$.
In terms of the new variables
\begin{equation}
\left\{
\begin{array}{lll}
Q_1&\equiv&\displaystyle i\partial_1+{g \nu^a \over 2} H x_2 \\
Q_2&\equiv&\displaystyle i\partial_2-{g \nu^a \over 2} H x_1
\end{array}
\right.
\end{equation}
we can write
\begin{equation}
\Delta^a = \nabla^2_{{\bf S}^2} - Q_1^2 -Q_2^2
\end{equation}
What is more, given that
$\displaystyle [ Q_1, Q_2 ]= -i g \nu^a H$,
one may take the representation
\begin{equation} 
Q_2= i g \nu^a H {\partial \over \partial Q_1} 
\end{equation}
and put
\begin{equation}
\begin{array}{lll}
\Delta^a &=&\displaystyle
\nabla^2_{{\bf S}^2}
- Q_1^2 + (g \nu^a H)^2 {\partial^2 \over \partial Q_1^2} \\
&=&\displaystyle \nabla^2_{{\bf S}^2}-2 {\cal H}_{\omega^a} ,
\end{array}
\end{equation}
where ${\cal H}_{\omega^a}$ indicates the QM-Hamiltonian
of a harmonic oscillator with frequency
\begin{equation}
\omega^a\equiv gH | \nu^a | .
\label{defomegaa}
\end{equation}
Let us now diagonalize the spacetime-index structure of
\begin{equation}
\overline{\Theta}_{\mu \nu}^a = -g_{\mu\nu} \Delta^a
+ig \nu^a \overline{F}_{\mu \nu} - R_{\mu \nu} .
\end{equation}
Taking into account that
\begin{equation}
\begin{array}{c}
g_{\mu \nu}= \mbox{diag}(g_{\theta \theta},g_{\varphi \varphi}, 1, 1 ),
\\
R_{\mu \nu}=\left\{
\begin{array}{ll}
\displaystyle{1 \over \rho^2} g_{\mu \nu},&
\mbox{when $\mu,\nu$ are ${\bf S}^2$ indices} \\
0,&
\mbox{when $\mu,\nu$ are ${\bf R}^2$ indices}
\end{array}
\right.
\end{array}
\end{equation}
($\rho$ is the radius of  ${\bf S}^2$)
and the particular choice of $\overline{F}_{\mu \nu}$,
we write $\overline{\Theta}_{\mu \nu}^a$ in
spacetime-matrix form:
\begin{equation}
\left( \overline{\Theta}_{\mu \nu}^a \right)=
\left(
\begin{array}{cccc}
-g_{\theta \theta}( \Delta^a +1/\rho^2 )&0  &0           &0 \\
0&-g_{\varphi \varphi}( \Delta^a +1/\rho^2 )&0           &0 \\
0&0                                         &-\Delta^a   &2i g\nu^a H \\
0&0                                         &-2i g\nu^a H&-\Delta^a \\
\end{array}
\right)
\end{equation}
Its diagonalization produces again four `eigenoperators', which are
\begin{equation}
-g_{\theta \theta}( \Delta^a +1/\rho^2 ) ,  \hspace{2mm}
-g_{\varphi \varphi}( \Delta^a +1/\rho^2 ) ,  \hspace{5mm}
-\Delta^a \pm 2 g | \nu^a | H
\end{equation}
All of them can be recast into the common notation
\begin{equation}
\overline{\Theta}_{\mu}^a \equiv -g_{\mu\mu}(\Delta^a + m_{\mu}^a), \
\left\{
\begin{array}{c}
m_{\theta}^a= m_{\varphi}^a= 1/\rho^2, \\
m_1^a= - m_2^a= -2 \omega^a ,
\end{array}
\right.
\label{notms}
\end{equation}
(where (\ref{defomegaa}) has been recalled) and one just has to consider
the eigenvalues of $-(\Delta^a + m_{\mu}^a)$, which are
\begin{equation}
{1 \over \rho^2}l(l+1)+2\omega^a\left( n+ {1 \over 2} \right) -m_{\mu}^a
\equiv \Lambda_{\mu}^a(n,l),
\mbox{with degeneracy $d_l= 2l+1 $, $l \in {\bf N}, n \in {\bf N}$ }
\label{Lamuanl}
\end{equation}
Although not calling it this name,
zeta-function regularization was applied to problems of similar
nature as early as the time of ref. \cite{SS}
(for a review, see \cite{EORBZ}).
Now that the spectrum is known,
we construct the zeta function for the whole $\overline{\Theta}$ operator
\begin{equation}
\zeta_{\overline{\Theta}}(s) = {\rm Tr}\  \overline{\Theta}^{-s}
= \sum_a \sum_{\mu} \int d^4x \sqrt{g} \
\mbox{tr} \,
\langle x |  \left(\overline{\Theta}_{\mu}^a \right)^{-s}  | x \rangle
\end{equation}
where tr runs over the harmonic oscillator and angular momentum
quantum numbers (so that ${\rm Tr}\ $ runs over everything).
However, the above eigenvalues were referred to the $Q_1$- and
not the $x$- coordinates. Introducing the adequate spectral resolution of
the identity and bearing in mind the property
$\displaystyle
\int dQ_1 | \langle x | Q_1, n, l \rangle |^2 =
{\omega^a \over 2 \pi} ,
$
we find
\begin{equation}
\zeta_{\overline{\Theta}}(s)= {\Omega \over 2\pi (4 \pi \rho^2 )}
\sum_a \omega^a \sum_{\mu} \sum_n \sum_l d_l [ \Lambda_{\mu}^a(n,l) ]^{-s},
\end{equation}
where
$\Omega\equiv \int \sqrt{g} \, d^4 x = 4\pi\rho^2 \int dx_1 \, dx_2$
denotes the total spacetime
volume.
Using the form of $\Lambda_{\mu}^a(n,l)$ given by (\ref{Lamuanl}) and
the dimensionless variables
\begin{equation}
\widehat{\omega}^a \equiv \rho^2 {\omega}^a , \hspace{1cm}
\widehat{m}_{\mu}^a \equiv \rho^2 m_{\mu}^a ,
\label{defhats}
\end{equation}
this zeta function is more conveniently expressed as
\begin{equation}
\zeta_{\overline{\Theta}}(s) =
{\Omega \over 8\pi^2 \rho^2 } \rho^{2s} \sum_a \omega^a \sum_{\mu}
\zeta(s; \widehat{\omega}^a, \widehat{m}_{\mu}^a) ,
\end{equation}
where
\begin{equation}
\zeta(s; \widehat{\omega}, \widehat{m} ) \equiv
\sum_{l=0}^{\infty} \sum_{n=0}^{\infty} (2l+1)
\left[
l(l+1) + 2\widehat{\omega} \left( n+{1 \over 2} \right) - \widehat{m}
\right]^{-s} .
\label{zeta3}
\end{equation}
Doing a Mellin transform, summing the $n$-series
afterwards, and making a trivial variable change (\ref{zeta3}),
can be put into the integral form
\begin{equation}
\zeta(s; \widehat{\omega}, \widehat{m})=
{1 \over (2\widehat{\omega})^s}
{1 \over \Gamma(s)} \int_0^{\infty} du \, u^{s-1} \,
{e^{\displaystyle -{u\over 2}
\left( 1 -{ \widehat{m} \over \widehat{\omega} } \right) }  \over
1-e^{-u} } \, Y_{{\bf\rm S}^2}\left( {u \over 2 \widehat{\omega}} \right).
\label{irzeta3}
\end{equation}
The advantage here is that the quantity
\begin{equation}
Y_{{\bf\rm S}^2}(t)\equiv \sum_{l=0}^{\infty} (2l+1) e^{ -t \, l(l+1) }
\label{hks2}
\end{equation}
is in fact the integrated heat-kernel
of the ($-$Laplacian) operator on an ${\bf S}^2$ sphere of unit radius
($t$ has absorbed the radius dependence, as this variable
stands for
$\displaystyle {u \over 2 \widehat{\omega}}= {u \over 2 \rho^2\omega}$).

The  $-\overline{D}^2$ operator is similar but for all the
$m_{\mu}^a$ pieces additional to $\Delta^a$ which are now absent.
Since the spacetime structure is already
diagonal, we do not have to take the final
$\displaystyle\sum_{\mu}$ for these
eigenvalues, and one can straightforwardly put
\begin{equation}
\zeta_{-\overline{D}^2}(s) =
{\Omega \over 8\pi^2 \rho^2 } \rho^{2s} \sum_a \omega^a
\zeta(s; \widehat{\omega}^a, 0) .
\end{equation}

Next we go on to the dimensionless versions of both operators by
introducing the arbitrary mass scale $\mu$, and form the linear
combination (\ref{wholez}):
\begin{equation}
-{1 \over 2} \zeta_{\overline{\Theta}/\mu^2}(s) 
+ \zeta_{-\overline{D}^2/\mu^2}(s) =
{\Omega \over 8\pi^2 \rho^2 } (\rho\mu)^{2s} \sum_a \omega^a
\left[
-{1 \over 2} 
\sum_{\lambda} \zeta(s; \widehat{\omega}^a, \widehat{m}_{\lambda}^a)
+\zeta(s; \widehat{\omega}^a, 0)
\right] .
\label{wholez1}
\end{equation}
As already observed, the derivative of this function at $s=0$ yields
the effective action.
Further, the representation (\ref{irzeta3}) enables us to write
\begin{equation}
\begin{array}{c}
\displaystyle
-{1 \over 2} 
\sum_{\lambda} \zeta(s; \widehat{\omega}^a, \widehat{m}_{\lambda}^a)
+\zeta(s; \widehat{\omega}^a, 0)= \\
\displaystyle {1 \over (2\widehat{\omega}^a)^s}
{1 \over \Gamma(s)} \int_0^{\infty} du \, u^{s-1} \,
{e^{\displaystyle -{u\over 2}}  \over 1-e^{-u} } \,
\left( -{1 \over 2}\sum_{\lambda= \theta, \varphi, 1,2}
e^{\displaystyle {u \over 2}{ \widehat{m}_{\lambda}^a \over \widehat{\omega}^a } }
+ 1 \right) \,
Y_{{\bf\rm S}^2}\left( {u \over 2 \widehat{\omega}^a} \right) .
\end{array}
\label{irzt}
\end{equation}
Recalling (\ref{notms}) one realizes that
$\displaystyle
\sum_{\lambda= \theta, \varphi, 1,2}
e^{\displaystyle {u \over 2}
{ \widehat{m}_{\lambda}^a \over \widehat{\omega}^a } }
= 2 \, e^{\displaystyle {u \over 2 \widehat{\omega}^a } } +2 \cosh u .
$
Therefore, (\ref{wholez1}) reads
\begin{equation}
\begin{array}{c}
-{1 \over 2} \zeta_{\overline{\Theta}/\mu^2}(s) 
+ \zeta_{-\overline{D}^2/\mu^2}(s)
= \\
\displaystyle{\Omega \over 8\pi^2 } \sum_a
{ (\omega^a)^2 \over \widehat{\omega}^a }
\left( \mu^2 \over 2\omega^a \right)^s
{1 \over  \Gamma(s)}\int_0^{\infty} du \, u^{s-1} \,
{e^{\displaystyle -{u\over 2}} \over 1-e^{-u} } \,
\left( -e^{\displaystyle {u \over 2 \widehat{\omega}^a } } 
- \cosh u +1 \right) \,
Y_{{\bf\rm S}^2}\left( {u \over 2 \widehat{\omega}^a} \right) .
\end{array}
\label{wholez2}
\end{equation}

\subsection{Large-$\rho$ expansion}
As is well known, the small-$t$ asymptotic expansion for the heat-kernel
of a Laplacian-like operator on a two-dimensional manifold ${\cal M}$
without boundaries has the form
\begin{equation}
Y_{\cal M}(t) \sim
{1 \over 4\pi t} \left[ A_0 + A_1 t + O(t^2) \right] \equiv
{1 \over t} \left[ a_0 + a_1 t + O(t^2) \right].
\end{equation}
In particular, for the Laplacian on ${\cal M}=S^2$ the
heat-kernel $Y_{{\bf\rm S}^2}$ (\ref{hks2}) gives rise to
the specific values \cite{MKS}
\begin{equation}
a_0=1, a_1=-{1 \over 3}, \dots
\label{aks}
\end{equation}
Since $\widehat{\omega}^a= \rho^2 \omega^a = \rho^2 gH | \nu^a |$,
expansions in $1/\widehat{\omega}$ are actually expansions in $1/\rho^2$.
Thus, calling $\displaystyle{u \over 2 \widehat{\omega}^a}\equiv t$, 
one finds
\begin{equation}
( -e^t -\cosh u +1 ) \, Y_{{\bf\rm S}^2}(t)
\sim -\left[
a_0 {1 \over t} \cosh u
+a_0 +a_1 \cosh u + O(t)
\right]
\label{smtexp}
\end{equation}
Putting this series into (\ref{wholez2}), we integrate term by term
with the help of the Hurwitz zeta function
integral representation
\begin{equation}
\zeta_H(z, \alpha)={1 \over \Gamma(z)}\int_0^{\infty} du \,
u^{z-1} \, {e^{-\alpha u} \over 1-e^{-u}}, \ {\rm Re}\  z > 1 .
\end{equation}
Substituting afterwards the result into (\ref{wholez2}) one gets
\begin{equation}
\begin{array}{c}
\displaystyle -{1 \over 2} \zeta_{\overline{\Theta}/\mu^2}(s) 
+ \zeta_{-\overline{D}^2/\mu^2}(s) =
\\
\begin{array}{ll}
\displaystyle -{\Omega \over 8\pi^2} \sum_a
2 (\omega^a)^2 \left( \mu^2 \over 2\omega^a \right)^s &
\displaystyle\left\{
{a_0 \over 2} {1 \over s-1} \left[
\zeta_H\left( s-1, {3 \over 2} \right)
+ \zeta_H\left( s-1, -{1 \over 2} \right)
\right] \right. \\
&\hspace{-10mm} \displaystyle\left. +\left[
a_0 \zeta_H\left( s, {1 \over 2} \right)
+{a_1 \over 2} \left(
\zeta_H\left( s, {3 \over 2} \right)
+\zeta_H\left( s, -{1 \over 2} \right)
\right)
\right]
{1 \over 2\widehat{\omega}^a }
+O\left(  1 \over (2\widehat{\omega}^a)^2 \right)
\right\}
\end{array}
\end{array}
\end{equation}
It will now be easy to take the derivative at $s=0$. When doing so, we
shall recall the identities
\begin{equation}
\begin{array}{lll}
\zeta_H(0,x)&=&\displaystyle -B_1(x)={1 \over 2}-x, \\
\zeta_H'(0,x)&=&\displaystyle -{1 \over 2}\ln(2 \pi ) + \ln\Gamma(x), \\
\zeta_H(-1,x)&=&\displaystyle -{1 \over 2}B_2(x)=-{1 \over 2}(x^2-x+1/6),
\end{array}
\label{zzp0m1}
\end{equation}
and observe that imaginary parts emerge from
\begin{equation}
\left\{
\begin{array}{lll}
\displaystyle\zeta_H'\left( s , -{1 \over 2} \right)&=&
\displaystyle\zeta_H'\left( s , {1 \over 2} \right)
-\left( -{1 \over 2} \right)^{-s} \ln\left( -{ 1 \over 2 } \right), \\
\displaystyle\ln\left( -{ 1 \over 2} \right)&=&
\displaystyle\ln\left( { 1 \over 2} \right) -i \pi ,
\end{array}
\right.
\end{equation}
where the complex argument determination taken is
that from refs. \cite{ES}.
Thus
\begin{equation}
\begin{array}{c}
\Gamma^{(1)} =
\displaystyle -{1 \over 2} \zeta_{\overline{\Theta}/\mu^2}'(0) 
+ \zeta_{-\overline{D}^2/\mu^2}'(0) =
\\
\begin{array}{ll}
\displaystyle
-{\Omega\over 8\pi^2}\sum_{a=1}^{N^2-1}2 (\omega^a)^2&
\displaystyle\left\{
{a_0\over 2} {11 \over 12} \ln\left( \mu^2 \over 2 \omega^a \right)
+{a_0\over 2} \left[ {11 \over 12}
- \zeta_H'\left( -1 , {3 \over 2} \right)
- \zeta_H'\left( -1 , {1 \over 2} \right)
+{1 \over 2}\ln\left( 1 \over 2 \right)
+i {\pi \over 2}
\right]  \right. \\
&\displaystyle\left. +{1 \over 2\widehat{\omega}^a }\left[
-{(a_0+a_1) \over 2} \ln 2 + i {a_1\over 2}\pi
\right]
+O\left( 1 \over ( 2\widehat{\omega}^a )^2 \right)
\right\} .
\end{array}
\end{array}
\label{e40}
\end{equation}
While the leading part for $\rho\to\infty$ amounts to the result in
\cite{ES}
---which is a generalization of Savvidy's  one \cite{Sa}--- 
the next-to-leading
contribution provides the first curvature correction and is, in fact,
linear in the curvature.
Let us now consider the particular case of $N=2$ in this
approximation. For SU(2),
the adjoint representation yields
$\{ |\nu^a| , a=1, \dots, N^2-1 \} = \{ 1,1,0 \}$.
As a result of this, and recalling (\ref{defomegaa}), (\ref{defhats}),
\begin{equation}
{\Gamma^{(1)} \over \Omega} =
a_0 (gH)^2\left[
{11 \over 48 \pi^2}
\left( \ln{ gH \over {\mu'}^2 } -{1 \over 2} \right)
-i {1 \over 8 \pi  }
\right]
     +{1 \over 4\pi^2}{gH\over\rho^2}
\left[
-{(a_0+a_1) \over 2} \ln 2 + i {a_1\over 2}\pi
\right] ,
\end{equation}
where the new arbitrary scale $\mu'$ is related to the initial arbitrary
scale $\mu$ by the transformation
$$
\ln {\mu'}^2 = \ln \mu^2 -{5 \over 11}\ln 2 + {1 \over 2}
-{11 \over 12}\left[
- \zeta_H'\left( -1 , {3 \over 2} \right)
- \zeta_H'\left( -1 , {1 \over 2} \right)
\right] .
$$
>From here on we will rename $\mu'$ into $\mu$. Next, we turn to the
1-loop effective potential $V$ including the classical contribution,
namely
\begin{equation}
V= {H^2 \over 2}+{\Gamma^{(1)} \over \Omega} ,
\end{equation}
whose real and imaginary parts read, after setting the values
(\ref{aks}), are given by
\begin{equation}
\begin{array}{lll}
\displaystyle{\rm Re}\  V(H,\rho)&=&\displaystyle {H^2 \over 2}
+
{11 \over 48 \pi^2} (gH)^2
\left( \ln{ gH \over {\mu }^2 } -{1 \over 2} \right)
+{1 \over 4\pi^2}{gH\over\rho^2}
{1 \over 3}
\ln 2 \\
\displaystyle{\rm Im}\  V(H,\rho)&=&\displaystyle
{1 \over 8 \pi  }\left[
-
(gH)^2
+{1 \over 3}
{gH\over\rho^2} \right] .
\end{array}
\end{equation}
They are plotted in Fig. 1 in solid and dashed line, respectively,
for several values of $\rho$, including the
flat space case corresponding to $\rho\to\infty$, in the order
---from bottom up:
 flat space, $\rho=10, 3, 2.21, 2.03, 1.5, 1$.
For this plot we have set $g=0.1$ and
the arbitrary $\mu^2$ scale has been chosen to be
$\displaystyle \mu^2= g e^{24\pi^2/(11g^2)}$ 
so that the flat space minimum
of ${\rm Re}\  V$ occurs at $H=1$.
Thus, we can see how the ${\rm Re}\  V$ minima are shifted to lower
$H$-values as $\rho$ decreases.
Moreover, a curvature-induced symmetry-breaking phase transition takes
place
at $\rho_c\simeq 2.03$: for $\rho$ larger than this value, the global
minimum is located at some $H_{\rm min} > 0$;
otherwise $H_{\rm min} = 0$.
We also observe that the zero of the imaginary part and the minimum of
the real part are in general different, but at
$\rho_s\simeq 2.21$ they coincide.
This indicates a possible stabilization of the magnetic
vacuum due to gravitation.
In the fourth section we shall calculate RG improved
effective potential and find also the possibility of curvature-induced
phase transitions.

\section{Gauge field effective potential on a
${\bf T}^2 \times {\bf R}^2$ background}

Let the radii of the torus be $L_1$ and $L_2$. For the rest
we shall use the same notation as in sect. 2, but this time the
curvature will be taken to be zero. As a result,
for the eigenvalues of $-(\Delta^a + m_{\mu}^a)$, we have now
(instead of (\ref{Lamuanl}))
\begin{equation}
\Lambda_\mu^a (n_1,n_2,n) =
\left({n_1 \over L_1}\right)^2+\left({n_2 \over L_2}\right)^2
+2\omega^a\left( n+ {1 \over 2}
 \right) -m_{\mu}^a, \qquad
n_1,n_2 \in \mbox{\bf Z}, n \in \mbox{\bf N} ,
\label{Lamuanx1}
\end{equation}
where
\begin{equation}
m_{t1}^a=  m_{t2}^a= 0, \ \
m_1^a= - m_2^a= -2 \omega^a .
\label{notmx1}
\end{equation}
We can proceed again by writing the corresponding zeta function
\begin{equation}
\zeta_{\overline{\Theta}}(s)= {\Omega \over 2\pi L_1 L_2}
\sum_a \omega^a \sum_{\mu} \sum_n \sum_{n_1,n_2}
\Lambda_{\mu}^a(n_1,n_2,n)^{-s},
\end{equation}
with $\Omega$ the total spacetime volume.
Introducing  $\bar{L}_i^a \equiv L_i \sqrt{2 \omega^a}$, $i=1,2$,
$\bar{m}_\mu \equiv m_\mu^a/(2\omega^a)$, we can write
\begin{equation}
\zeta_{\overline{\Theta}}(s) =
{\Omega \over 4\pi L_1L_2 }  \sum_a (2\omega^a)^{1-s} \sum_{\mu}
\zeta(s; \bar{L}_i^a, \bar{m}_{\mu}) ,
\end{equation}
where
\begin{equation}
\zeta(s; \bar{L}_i^a, \bar{m}_{\mu})  \equiv
\sum_{n=0}^{\infty} \sum_{n_1,n_2= -\infty}^{+\infty}
\left[
\left({n_1 \over \bar{L}_1}\right)^2+\left({n_2 \over \bar{L}_2}\right)^2
+ n+ \alpha_\mu
\right]^{-s} ,
\label{zetax3}
\end{equation}
being $\alpha_\mu \equiv 1/2 - \bar{m}_\mu$.

After some work that involves
performing, as before, the usual Mellin transform, using
later Jacobi's theta function identity on the sums over the indexes
$n_1$ and $n_2$, and integrating back the the Mellin transform (see
\cite{EORBZ}), we obtain
\begin{eqnarray}
&& \hspace{-4mm} -{1 \over 2}
\sum_{\lambda} \zeta(s; L_i, \alpha_\lambda)
+\zeta(s; L_i, 0) \ = \
\frac{\Omega}{2\pi L_1L_2}\sum_{a,\lambda} \omega^a \left( \frac{\mu^2}{
2\omega^a} \right)^s \left[2 \pi L_1 L_2 \omega^a \zeta (s-1;
 \alpha_\lambda) -\zeta(s;\alpha_\lambda ) \right. \nonumber \\
&& + \left. \frac{2\pi^s L_1 L_2 (2\omega^a)^{(s+1)/2}}{\Gamma (s)}
\sum_{n,n_1,n_2} \left( \frac{L_1^2n_1^2 + L_2^2 n_2^2}{n+\alpha_\lambda}
\right)^{(s-1)/2} K_{s-1} \left( 2\pi \sqrt{2\omega^a (n+\alpha_\lambda)
(L_1^2n_1^2+L_2^2n_2^2)} \right) \right]. \nonumber \\
\label{irztx1}
\end{eqnarray}

Finally, by taking the derivative of this zeta function,
for the one-loop effective action density we arrive at the following
expression
\begin{eqnarray}
\hspace{-3mm}  {\Gamma^{(1)} \over \Omega} (L_1,L_2) &=&
\sum_{a=1}^{N^2-1} \left\{ (\omega^a)^2 \left[ -
 {11 \over 12} \ln\left( \mu^2 \over 2 \omega^a \right)
+ {11 \over 12}
- \zeta_H'\left( -1 , {3 \over 2} \right)
- \zeta_H'\left( -1 , {1 \over 2} \right)
+{1 \over 2}\ln\left( 1 \over 2 \right)
+i {\pi \over 2} \right]  \right. \nonumber \\
&& +   {\omega^a \over 2\pi L_1L_2}  ( \ln 2 - i \pi)
+ {\sqrt{2} \over \pi} (\omega^a)^{3/2}  
\label{irztx2} \\
&& \left. \cdot \sum_\mu \sum_{n,n_1,n_2}
\sqrt{\frac{n+\alpha_\mu}{L_1^2n_1^2+L_2^2n_2^2}} \ K_1
 \left( 2\pi \sqrt{2\omega^a (n+\alpha_\lambda)
(L_1^2n_1^2+L_2^2n_2^2)} \right) \right\}.
\nonumber
\end{eqnarray}
The difference from Eq. (\ref{e40}) is that the present
effective action is explicitly given in terms of
 a quickly (exponentially like)
convergent series, as far as $L_1,L_2 >> 1$. For the first terms of the
expansion we get a qualitative coincidence with the large-$\rho$ expansion
of the preceding case. This allows us to stop the discussion of the torus case
here. In particular, the numerical analysis and that of the imaginary part
closely parallel the one carried out in the previous section.

\section{RG-improved gauge field effective potential in an SU(2)
gauge theory in curved spacetime}
Let us consider now the SU(2) gauge theory with the Lagrangian
\begin{equation}
{\cal L}=-{1 \over 4}G_{\mu\nu}^a G^{\mu\nu \ a} +a_1 R^2 +a_3 G
\end{equation}
on the constant-curvature background
$R_{\mu\nu}={1 \over 4}R g_{\mu\nu}$, where the curvature is small
enough. Such a theory is multiplicatively renormalizable in curved
spacetime (for details, see \cite{BOS}). Here $G$ is the Gauss-Bonnet
combination, and $a_1$, $a_3$ are the usual gravitational coupling
constants.

The covariantly constant gauge field choice may be understood as
an expansion in normal coordinates. Then, the first term in this
expansion gives the flat space result and the next the curvature
corrections:
$\displaystyle 
{1 \over 4}G_{\mu\nu}^a G^{\mu\nu \ a}={1 \over 2}H^2 + O(R)$.

Taking into account that the theory under consideration is
multiplicatively renormalizable, and solving its renormalization group
equations for the effective potential we find (see also \cite{OP})
\begin{equation}
V={1 \over 2} {g^2 \over g^2(t)} H^2 -a_1(t)R^2-a_3(t)G,
\end{equation}
where $G={R^2 \over 6}$ and
\begin{equation}
g^2(t)={g^2 \over \displaystyle 1+{11 g^2 t \over 12 \pi^2}}, \hspace{5mm}
a_1(t)=a_1, \hspace{5mm}
a_3(t)=a_3-{62 t \over 120 (4\pi)^2}.
\label{RGV}
\end{equation}
Hence, we have got the RG-improved effective potential making the
summation of all the leading logs of perturbation theory. The RG
parameter is chosen as the logarithm of the effective gauge mass
(compare with (\ref{defTheta}))
\begin{equation}
t={1 \over 2} \ln { { {R\over 4} +gH } \over \mu^2} .
\end{equation}
Such a choice may also be justified by a direct one-loop calculation
near
flat-space, and then choosing the standard condition on the vanishing
of the logarithmic term in the one-loop effective potential. Note also
that from (\ref{RGV}) at $R=0$ we get the RG-improved effective potential
of paper \cite{Sa}.

Now one can find the minimum of the effective potential, which turns out
to be (see also \cite{OP})
\begin{equation}
g H_{\rm min} \simeq
\mu^2 \exp\left( -{24 \pi^2 \over 11 g^2} \right) - {R \over 4} .
\label{RGHmin}
\end{equation}
This vacuum takes into account the gravitational corrections. Moreover,
there is a critical value of the curvature determined by
$H_{\rm min}=0$ in (\ref{RGHmin}). Hence, like in third section
we see again the possibility of transitions between different
phases of the chromomagnetic vacuum. For positive curvatures above
the critical value $R_c$, the ground state is given by
$H_{\rm min}=0$, while for positive and small negative curvatures
below $R_c$, the ground state has a nonzero background gauge field
$H_{\rm min}\neq 0$. Note, however, that in the above quasi-local
expansion we cannot calculate the imaginary and nonlocal parts of the
effective action. Hence, it is not clear whether the above
$H_{\rm min}$ value is the true ground state or just a local minimum as
in flat space.

\section{Conclusion}
In the present letter we have discussed the gauge field effective
potential for a covariantly constant gauge field in curved spacetime.
The calculation has been done in two ways: explicit one-loop evaluation
on an ${\bf S}^2 \times {\bf R}^2$ background using zeta-regularization,
and RG-improved formalism using a quasi-local expansion. In
both approaches the influence of curvature on the phase structure is
clearly seen through the shift of the chromomagnetic vacuum due to
curvature corrections (curvature-induced phase transitions).
Moreover, numerical estimates on the ${\bf S}^2 \times {\bf R}^2$
background indicate that for some curvature value the nonzero
chromomagnetic vacuum may be stabilized (true ground state). Of course,
in order to completely confirm such a gravitational effect on the
chromomagnetic vacuum, explicit one-loop calculations on other
gravitational backgrounds should be done. Note, however, that for the
most interesting spacetimes (like De Sitter space ${\bf S}^4$), where
nonzero components of the background field lie in nonflat spacetime
directions, the spectrum gets extremely complicated and the one-loop
effective potential becomes really hard to calculate.

It would also be very interesting to study cosmological applications of
the above effective potential, for example, trying to construct an
inflationary Universe where $H$ plays the role of scalar inflaton.

\vskip5ex
\noindent{\Large \bf Acknowledgements}

We would like to thank J. Ambj{\o}rn and R. Percacci for
useful discussions.
This work has been supported by DGICYT (Spain), project
PB93-0035 and grant SAB93-0024, by CIRIT (Generalitat de
Catalunya), grant GR94-8001, and by RFBR, project 96-02-16017.

\newpage
\noindent{\Large \bf Figure Caption}
\vskip3ex
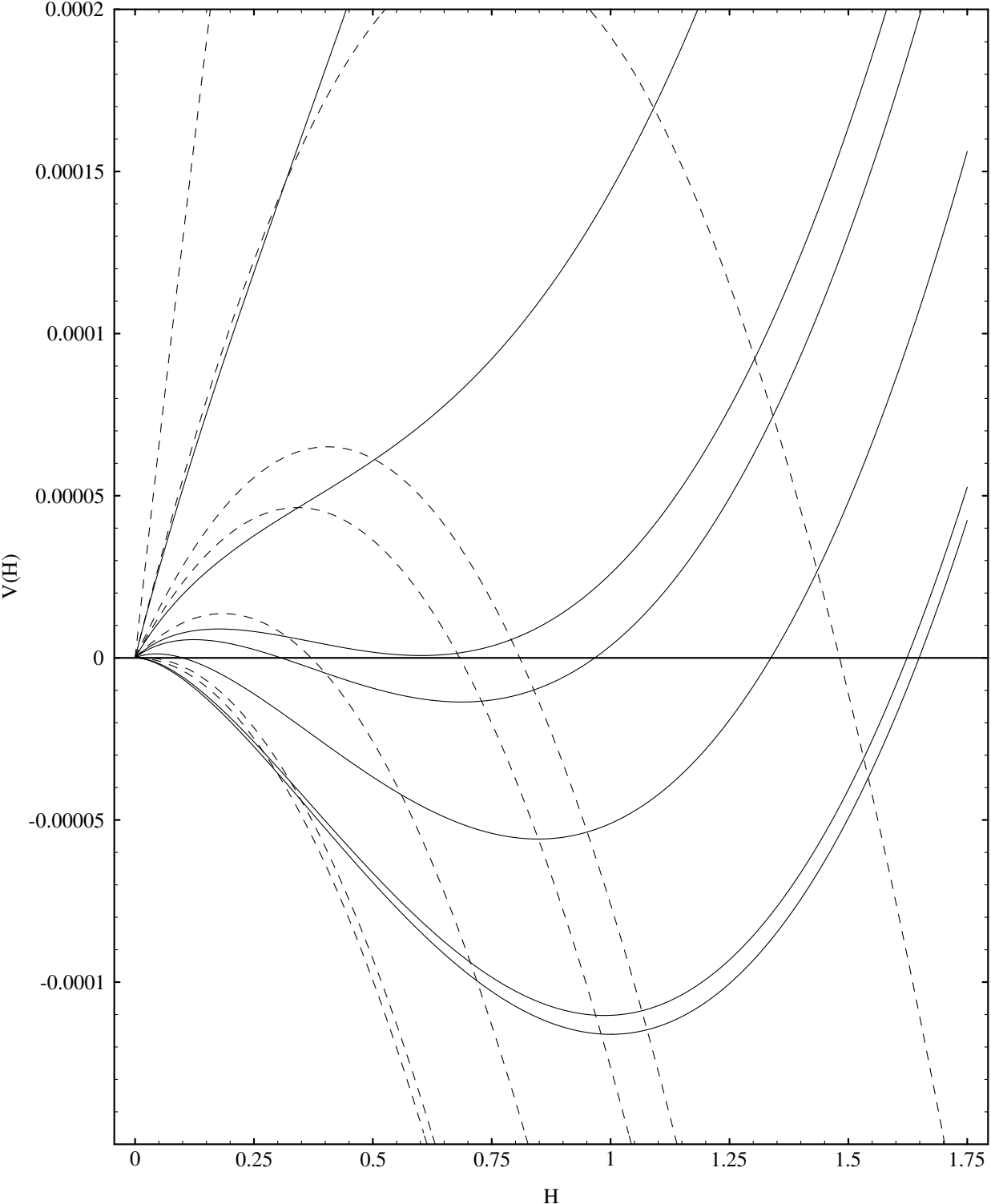
\noindent{\bf Fig. 1}.
 Re $V$ (solid lines) and  Im $V$ (dashed lines),
for several values of $\rho$.
>From bottom up, the curves correspond to
$\rho\to\infty$, $\rho=10, 3, 2.21, 2.03, 1.5, 1$ .
The value of the gauge coupling is $g=0.1$ and
the $\mu^2$ scale has been taken
$\displaystyle \mu^2= g e^{24\pi^2/(11g^2)}$ so that the minimum of Re $V$
for $\rho\to\infty$ takes place at $H=1$.

\end{document}